# Spatiotemporal Hologram

Qian Cao[1,2,3,†], Nianjia Zhang[1,†], Andy Chong[4,5], and Qiwen Zhan[1,2,3,6*]

[1] School of Optical-Electrical and Computer Engineering, University of Shanghai for Science and Technology, 200093 Shanghai, China
[2] Zhangjiang Laboratory, Shanghai, China
[3] University of Shanghai for Science and Technology, Shanghai Key Laboratory of Modern Optical System, Shanghai, China
[4] Department of Physics, Pusan National University, Busan, 46241, Republic of Korea
[5] Institute for Future Earth, Pusan National University, Busan, 46241, Republic of Korea
[6] Westlake Institute for Optoelectronics, Fuyang, Hangzhou 311421, China.
[†] These authors contribute equally to this work.
[*] Corresponding author: qwzhan@usst.edu.cn

**Spatiotemporal structured light has opened up new avenues for optics and photonics. Current spatiotemporal manipulation of light mostly relies on phase-only devices such as liquid crystal spatial light modulator to generate spatiotemporal optical fields with unique photonic properties. However, simultaneous manipulation of both amplitude and phase of the complex field for the spatiotemporal light is still lacking, limiting the diversity and richness of achievable photonic properties. In this work, a simple and versatile spatiotemporal holographic method that can arbitrarily sculpture the spatiotemporal light is presented. The capabilities of this simple yet powerful method are demonstrated through the generation of fundamental and higher-order spatiotemporal Bessel wavepacket, spatiotemporal crystal-like and quasi-crystal-like structures, and spatiotemporal flat-top wavepackets. Fully customizable spatiotemporal wavepackets will find broader application in investigating the dynamics of spatiotemporal fields and interactions between ultrafast spatiotemporal pulses and matters, unveiling previously hidden light-matter interactions and unlocking breakthroughs in photonics and beyond.**

## Introduction

With the rapid development of ultrafast lasers, the ability to sculpt spatiotemporal wavepackets has become increasingly important for both unravelling fundamental physics and fulfilling demanding industrial applications[1, 2]. Fully customizable spatiotemporal light holds immense potential in diverse fields such as laser machining[3], spatiotemporal edge detection[4], nonlinear optics[5, 6], optical fiber propagation[7], quantum optics[8], and various light-matter interaction studies[9, 10]. Traditionally, manipulating optical light fields has involved separate interventions in the spatial and temporal domains. For spatial light, holographic methods and programmable devices like liquid crystal spatial light modulators (LC-SLMs) have become prevalent tools for tailoring laser modes[11, 12], generating structured beams like nondiffracting beams[13], and even achieving arbitrary spatial profiles[14]. Pulse shaping in the time domain, on the other hand, typically occurs in the spectral domain using a 4-f setup and involves strategically placing diffractive elements or 1D LC-SLMs in the focal plane[15, 16].

In 2020, the demonstration of photonic cyclones carrying transverse orbital angular momentum (OAM) opened a new avenue for manipulating spatiotemporal structured light[17]. Utilizing a 2D spatial light modulator (SLM) within a conventional pulse shaper, scientists have spearheaded the study of spatiotemporally coupled light fields, revealing their significant physical implications and promising applications. These include space-time wavepackets with intricate spatiotemporal correlations[18, 19], spatiotemporal optical vortices with transverse OAM[17], and toroidal vortices of light[20].

Researchers recently developed a 3D wavepacket generation method combining pulse shaping with multi-plane light conversion (MPLC)[21]. While this technology enables multi-dimensional light sculpting, it struggles with producing tight spatiotemporal structures like optical vortices due to its approach of modulating the spatial light profile at each time delay. However, relying solely on a conventional pulse shaper consisting of a grating, a cylindrical lens, and a 2D SLM as a spatiotemporal phase modulator presents a significant obstacle to efficiently and precisely manipulating spatiotemporal wavepackets on demand.

This paper introduces a novel spatiotemporal hologram method for spatiotemporal complex amplitude modulation, enabling the custom shaping of arbitrary 2D spatiotemporal optical fields. We achieve this by incorporating a computer-generated hologram (CGH) into a conventional pulse shaper. The capabilities of this device are showcased through the generation of diverse and unprecedented spatiotemporal light structures, including fundamental and higher-order spatiotemporal Bessel wavepackets, spatiotemporal crystal-like and quasi-crystal-like structures, and spatiotemporal top-hat wavepackets.

**Results and Discussions**

Figure 1(a) illustrates the schematic of the spatiotemporal (ST) holographic setup and the associated interferometric characterization system. The input mode-locked laser centered at 1030 nm is divided into two parts by the beam splitter (BS1). One part is compressed by a pair of gratings, serving as the probe pulse. The other one is modulated into the desired wavepacket using the ST complex amplitude modulator. The probe pulse and the generated wavepacket are recombined at the CCD camera with a small tilted angle. By changing their relative time delay, the three-dimensional information of the generated wavepacket is recorded and later reconstructed[22].

The ST holographic setup consists a grating, a cylindrical lens and a phase-only LC-SLM. They are separated by the focal length of the cylindrical lens, forming a 4-f system. With such configuration, LC-SLM encoded with properly designed digital hologram can simultaneously modulate the spatial-spectral phase $\varphi(x,\omega)$ and amplitude $A(x,\omega)$ of the incoming optical field. The encoding technique utilizes a one-dimensional blazed phase grating with different modulation depth in order to diffract light into the first order with different light intensity[23]. Here, we use the technique conversely so that undesired light is diffracted into the first order while the desired light is preserved in the zeroth order. In order to achieve it, we place a small iris after the ST complex amplitude modulator so that high order diffracted light is fully blocked by the iris. Figure 1(b) shows the phase patterns of such blazed phase grating when the modulation depth $\phi_0 = 0.8\pi, 1.2\pi, 2\pi$, respectively. Figure 1(c) shows the theoretical energy distribution from a 1D blazed phase grating when $\phi_0$ varies from 0 to $2\pi$. Figure 1 (d) shows both theoretical and experimental intensity of the zeroth order light when $\phi_0$ varies. The experimental data obtained by the setup shown in Fig. 1(a) agrees well with the theoretical curve, indicating the possibility of precise control of the ST complex optical field. The measured extinction ratio of the light intensity ($I_{max}/I_{min}$) is about 125:1 in this figure. It is noteworthy that the intensity modulation of the light field in the zeroth order also accompanies with an additional retardance phase that should be properly compensated. Their relationship is shown in Fig. 1(e). The cause of this retardance phase is due to the change of overall phase in the blazed phase grating.

The results we discussed above use blazed phase grating with spatially uniform modulation depth $\phi_0$. To achieve arbitrary ST complex amplitude modulation $\Psi(x,\omega) = A(x,\omega)\exp[i\varphi(x,\omega)]$, the digital hologram loaded on SLM is

$$\varphi_{SLM}(x,\omega) = \phi_0(x,\omega) \cdot \frac{Mod(x,\Lambda)}{\Lambda} + \phi_r(x,\omega) + \varphi(x,\omega). \tag{1}$$

In this expression, the first term represents the blazed phase grating with a period of $\Lambda$ and a modulation depth of $\phi_0(x,\omega)$. $\phi_0(x,\omega)$ is related with the amplitude modulation factor $A(x,\omega)$ via the relationship shown in Fig. 1(d); the second term $\phi_r(x,\omega)$ is the phase compensation term for the phase retardance; and the last term $\varphi(x,\omega)$ is applied spatial-spectral phase modulation.

This encoding technique for ST complex amplitude modulation using programmable phase element offers us a much wider range of applications as it is able to arbitrarily sculp ST optical fields. In the process, the subtle interplay between dispersion and diffraction also plays a crucial role in generating the ST optical field. The balance between diffraction and dispersion at the distance $L$ and group delay dispersion (GDD) can be written as[24]

$$E(X,T;L,GDD) = \frac{w_b \cdot w_\Omega}{L} e^{ik_0\left(L+\frac{X^2}{L}\right)} \iint E_0(x',\Omega') e^{i\frac{GDD \cdot w_\Omega^2}{2}\Omega'^2} e^{ik_0 w_b^2\left(\frac{1}{2L}-\frac{1}{2f}\right)x'^2}$$
$$e^{i\Omega'T \cdot w_\Omega} e^{-ik_0\frac{x'X}{L}w_b} dx' d\Omega', \tag{2}$$

where $w_\Omega$ and $w_b$ are the spectral and spatial width of the input field, respectively, $x'/\Omega'$ are normalized spatial and spectral coordinate, $f$ is the focal length from the applied phase. The balance requires $GDD \cdot w_\Omega^2 = k_0 w_b^2 \left(\frac{1}{L}-\frac{1}{f}\right)$. An extra focusing or defocusing phase can determine the sign of GDD. By defining $\cot\gamma = GDD \cdot w_\Omega^2 = k_0 w_b^2 \left(\frac{1}{L}-\frac{1}{f}\right)$ and using the following substitution $x = \frac{(X \cdot \cos\gamma)}{w_b} \cdot \left(1+\frac{L}{f-L}\right)$, $t = -T \cdot w_\Omega \cdot \sin\gamma$, Equation (2) can be simplified to the following expression

$$E(X,T;\cot\gamma) = C \iint E_0(x',\Omega') e^{i\frac{\Omega'^2+x'^2}{2}\cot\gamma} e^{-i\Omega' t \csc\gamma} e^{-ixx' \csc\gamma} dx' d\Omega', \tag{3}$$

where C is a field constant. Equation (3) is called fractional Fourier transform when $0 < \gamma <$

$\pi/2$ and becomes complete Fourier transform when $\gamma = \pi/2$.

To demonstrate the ST complex amplitude modulation capacity, we first use the device to generate Bessel spatiotemporal wavepackets (BeSTWP) and higher order Bessel spatiotemporal optical vortices (BeSTOV) wavepackets. Both of these wavepackets have an annular distribution in Fourier space. Figure 2(a1-a4) presents the generation of the fundamental BeSTWP based on an annular ring mask in the spatial-spectral domain. The applied phase mask is shown in Fig. 2(a1). The desired hologram is generated using a blazed grating to preserve the light field located on the ring mask while diffract the light field elsewhere. The spatial width $w_b$ and spectral width $w_\Omega$ of the ring are 90 μm and 1THz, respectively. At the propagating length $L = 1.25$m, the calculated $\gamma$ is around 87.7°, indicating the ST field is closed to full Fourier transform form. Figure 2(b-c) show higher order BeSTOV results by applying additional ST spiral phase. BeSTOVs carrying transverse OAM with topological charge from – 2 to +2 are generated and they are confirmed by the measurement.

The generation of the fundamental BeSTWP and BeSTOV have proved the capability of the ST holographic setup. As a two-dimensional localized optical fields, BeSTWP and BeSTOV wavepackets have radial symmetrical distribution. There also exists another kind of localized states called optical lattices with translational symmetry and rotational symmetry. Using the ST holographic technique, spatiotemporal optical lattice can be realized in the spatiotemporal domain by selecting proper pinhole patterns in the spatial-spectral domain. Figure 3(a) illustrates several configurations to generate ST crystal-like and quasi-crystal-like structures in simulation. Each pinhole is uniformly located on a circle and the number of pinholes $q$ represents $q$-fold rotational symmetry. When the number of pinholes $q = 4$ and $q = 6$, the spatiotemporal optical time crystals have translational symmetry shown in Fig. 3(a, c). Especially, the spatiotemporal optical quasi-time crystal with rotational symmetry is generated at $q = 5$ in Fig. 3(b). It is worthy of noting that the five-fold rotation is preserved even with a gradient phase shown in Fig. 3(d-e). The transverse spiral phase can be also preserved in the spatiotemporal domain together with the five-fold rotation symmetry, forming a ST quasi-crystal structure carrying transverse OAM.

Following the simulations shown in Figure 3, Figure 4 both presents experimental generation of periodic and aperiodic spatiotemporal crystal structured light field. Figure 4(a, c) plot the reconstructed spatiotemporal crystals with 4-fold rotational symmetry and 6-fold rotational symmetry, respectively. Figure 4(b) plots spatiotemporal quasi-crystals with 5-fold rotational symmetry. Figure 4(d, e) show the effects of transverse spiral phase on the spatiotemporal quasi-crystals. The extra transverse OAM results a zero intensity in the center of spatiotemporal quasi-crystals and a phase singularity in spatiotemporal domain. The spatiotemporal quasi-crystal structure with transverse OAM maintains 5-fold rotational symmetry as predicted in Figure 3.

Using the spatiotemporal holographic technique, we can dynamically generate two-dimensional spatiotemporal wavepackets of arbitrary shape. As discussed above, all results are products of diffraction and dispersion in dynamic equilibrium. However, especially in optical materials processing, a static and higher power beam shaping in more important. Hence, we use an extra lens and de-chirped phase on the SLM to eliminate diffraction and dispersion to realize a spatiotemporal flat-top wavepackets in full Fourier transform. Unlike flat-top beams in the spatial domain and the flat-top pulses only in the time domain, the spatiotemporal flat-top wavepackets have both homogenous intensity distribution in $x - t$ domain. Figure 5(a) shows the 3D intensity map and projections view of the reconstructed spatiotemporal flat-top pulses. Figure 5(b) plots the accumulated intensity in the X-direction and T-direction, respectively. The spatial width (full-width at half-maximum) is 1 mm and the temporal width (full-width at half-maximum) is around 2 ps. Just as flat-top beams and flat-top pulses play an important role in the field of optical processing and optical computing, spatiotemporal flat-top pulse with uniform intensity in both the $x - t$ domain synthesized with this spatiotemporal holographic technique have broad potential applications in ultrafast laser machining as well as generating uniformly distributed electron wavepackets.

## Conclusions

To conclude, we introduce a novel spatiotemporal hologram that enables control over the complex field (both amplitude and phase) of optical spatiotemporal wavepackets. Both theoretical and experimental results demonstrate the capability of this technique for generating complex spatiotemporal wavepackets with arbitrary amplitude-phase profiles. Examples include fundamental and higher-order spatiotemporal Bessel wavepackets, spatiotemporal crystal and quasi-crystal structures, and spatiotemporal flat-top wavepackets. This manipulation of spatiotemporal complex amplitude expands the range of achievable optical properties in spatiotemporal wavepackets and promises broad applications in ultrafast processing, ultrafast imaging, optical communication, and exploring novel light-matter interactions.

## Acknowledgements

We acknowledge financial support from National Natural Science Foundation of China (NSFC) [Grant Nos. 92050202 (Q.Z.) and 12104309 (Q.C.)], the Shanghai Science and Technology Committee [Grant No. 19060502500 (Q.Z.)], the Shanghai Sailing Program [Grant No. 21YF1431500 (Q.C.)], National Research Foundation of Korea (NRF) [Grant No. 2022R1A2C1091890 (A.C.)], and Learning & Academic research institution for Master's·PhD students, and Postdocs (LAMP) Program of the National Research Foundation of Korea(NRF) grant [Grant No. RS-2023-00301938 (A.C.)]. Q. Z. also acknowledges support by the Key Project of Westlake Institute for Optoelectronics (Grant No. 2023GD007).

Q.C. and Q.Z. proposed the original idea and initiated this project. Q.C and N.Z. designed and performed the experiments. Q.C., N.Z., and Q.Z. analyzed the data. Q.C., N.Z., A.C., and Q.Z. prepared the manuscript. Q.C. and Q.Z. supervised the project. All authors contributed to the discussion and writing of the manuscript.

## Data availability

All the data are available in the article from the corresponding authors upon reasonable request.

## Conflict of interest

The authors declare no competing interest.

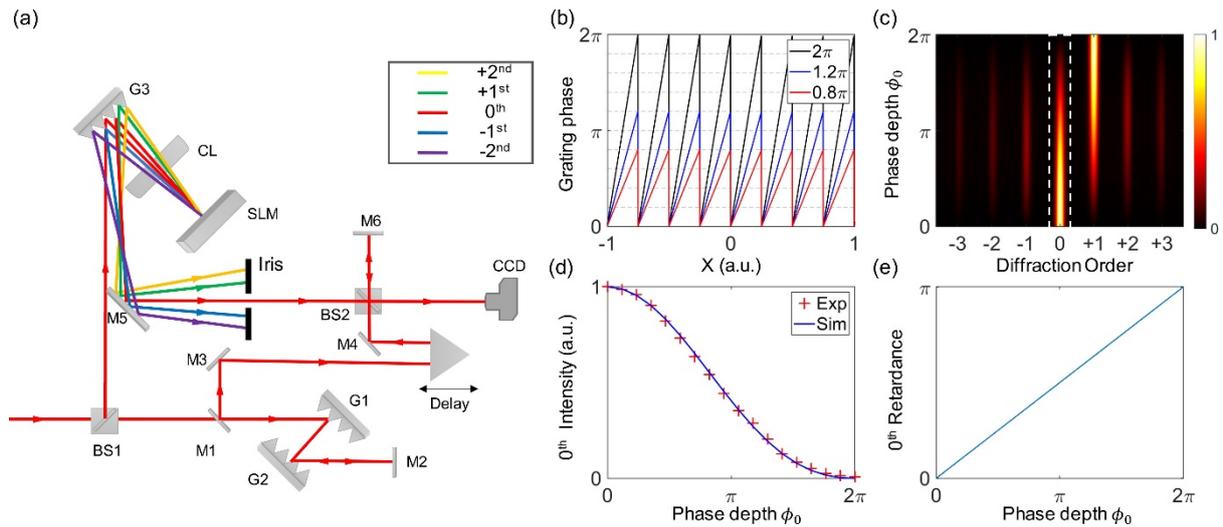

**Fig. 1 | Illustration for spatiotemporal hologram experimental setup and its working principal.**
(a) Realization of Spatiotemporal complex amplitude modulation with redistribution of energy into other diffraction order. BS: beam splitter, M: mirror, G: grating, CL: cylindrical lens. (b) One dimensional spatial linear grating with variable phase depth loaded onto the SLM. (c) Simulated process of redistribution of energy caused by variable modulated depth, resulting in the appearance of higher diffraction order and the variation of the $0^{th}$ order between 0 and 1. (d-e) Experimental results for the $0^{th}$ order intensity variations in Fig. 1(c) and simulation results for the $0^{th}$ order retarded phase caused by different modulation depth. The extinction ratio is around 125:1.

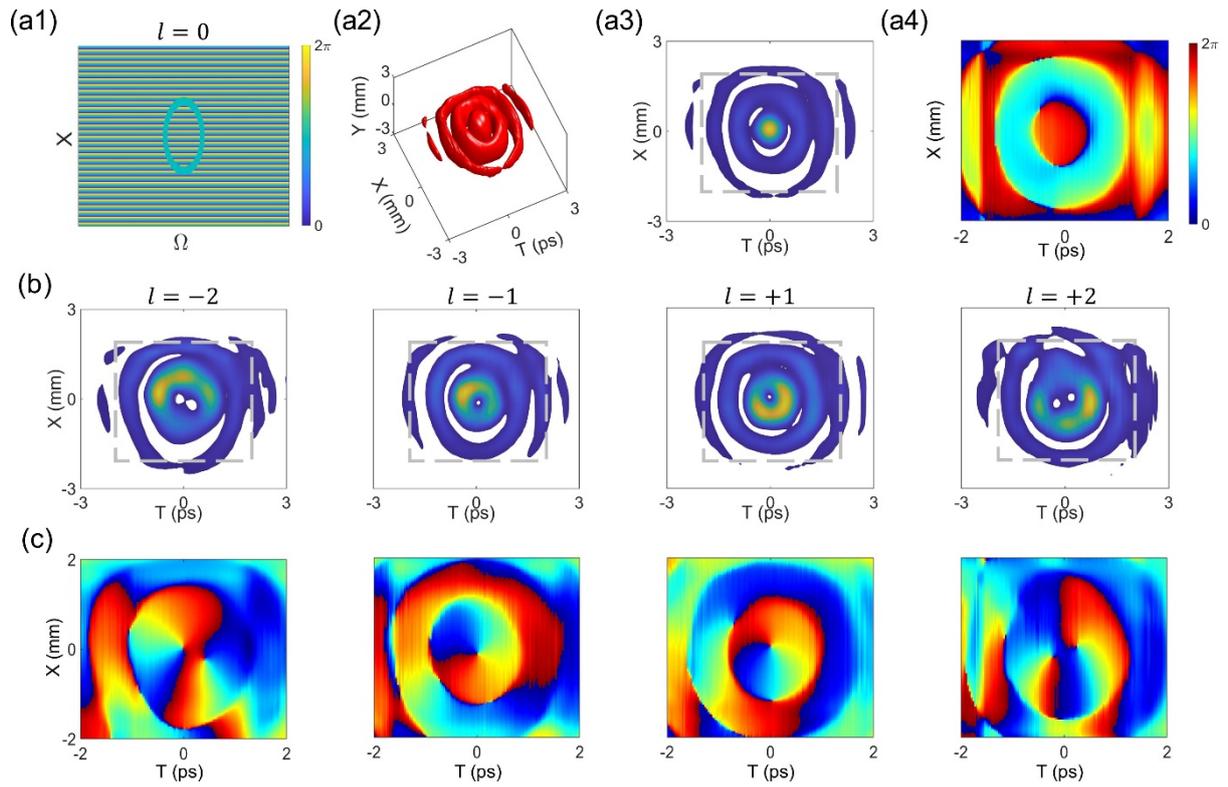

**Fig. 2 | Demonstration of spatiotemporal complex modulation with the generation of the family of BeSTWP.**
(a1) The phase mask of the fundamental BeSTWP; (a2-a4) 3D reconstructed intensity distribution with normalized iso-value 0.02 and 2D slice intensity and reconstructed phase. (b-c) spatiotemporal complex amplitude modulation for higher order (l from −2 to 2) BeSTWP with normalized iso-value 0.04.

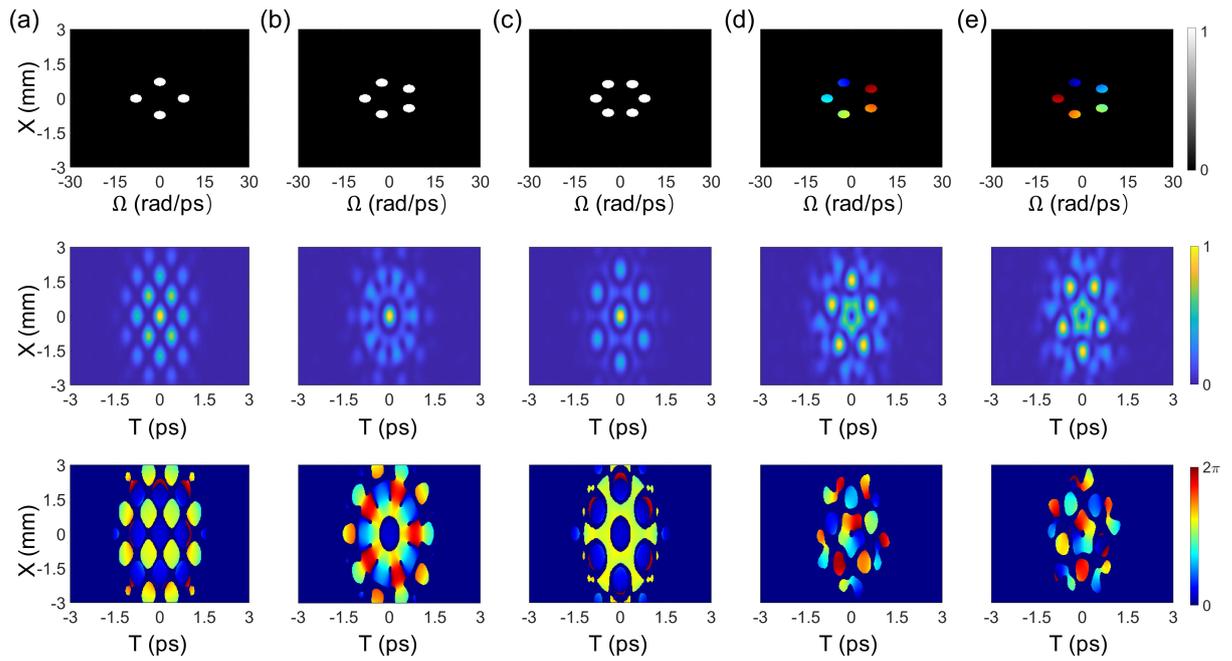

**Fig. 3 | Simulations of spatiotemporal optical time crystals and spatiotemporal optical time quasi-crystals.**
(a-c) the number of pinholes $q = 4$, $q = 5$ and $q = 6$ in spatial-spectral domain and its intensity and phase structure after Fourier transform. (d-e) the intensity and phase patterns in far field at $q = 5$ with the extra gradient transverse spiral phase $l$ from $-1$ to $+1$.

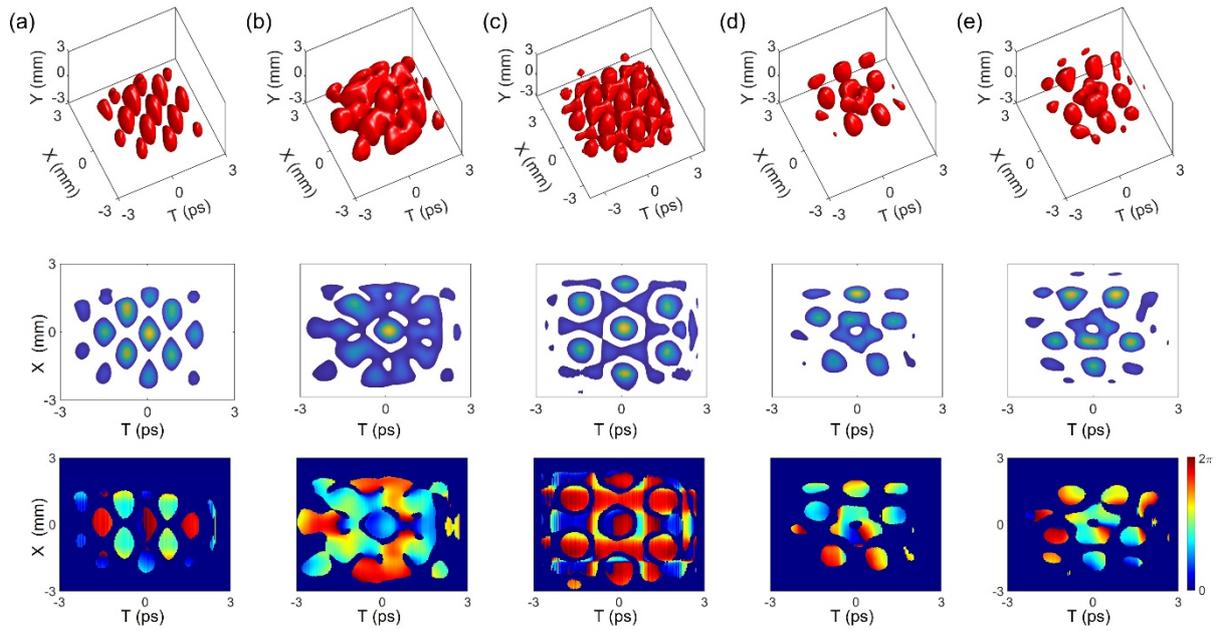

**Fig. 4 | Experimental results of spatiotemporal optical time crystals and spatiotemporal optical time quasi-crystals.**
(a-c) Reconstructions of spatiotemporal optical time crystals ($q = 4, 6$) and spatiotemporal optical time quasi-crystals ($q = 5$) with normalized iso-value 0.06. (d-e) Reconstructions of spatiotemporal optical time quasi-crystals with gradient transverse spiral phase from $-1$ to $1$ with normalized iso-value 0.1.

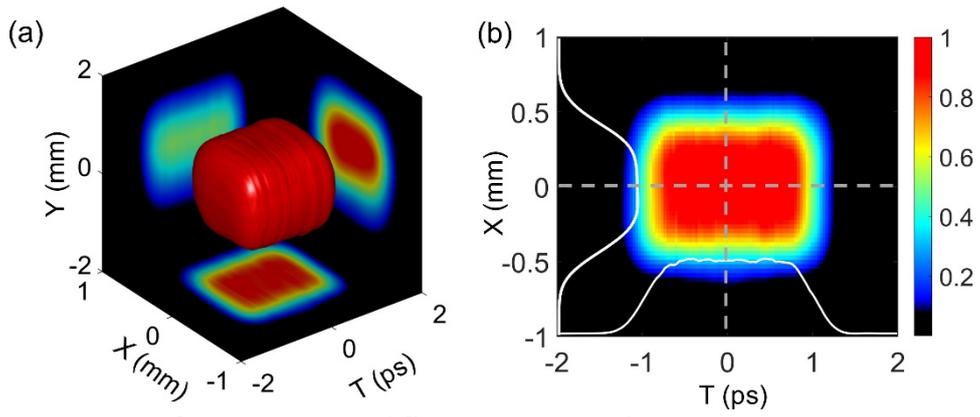

**Fig. 5 | Generation of spatiotemporal flat-top wavepacket.**
(a) 3D intensity isosurface of the spatiotemporal flat-top wavepacket with normalized iso-value 0.3 and its projections; (b) Accumulated intensity profile of spatiotemporal flat-top wavepackets and its cross lines along X = 0 and T = 0 axes.